\documentclass[letterpaper,conference]{IEEEtran}
\usepackage{amssymb}
\usepackage{verbatim}
\usepackage{graphicx}
\usepackage{amsfonts}
\usepackage{subfigure}
\usepackage[normalem]{ulem}
\usepackage{amsmath}
\usepackage{amsthm}
\usepackage[table]{xcolor}
\usepackage{mathtools}
\usepackage{epstopdf}

\begin{document}
\pagenumbering{gobble}

\title{Random Linear Fountain Code with Improved Decoding Success Probability}

\author{Jalaluddin Qureshi\\
Department of Electrical Engineering,\\
Namal College, Mianwali, Pakistan.\\
jala0001@e.ntu.edu.sg}

\maketitle

\begin{abstract}
In this paper we study the problem of increasing the decoding success probability of random linear fountain code over $GF(2)$ for small packet lengths used in delay-intolerant applications such as multimedia streaming. Such code over $GF(2)$ are attractive as they have lower decoding complexity than codes over larger field size, but suffer from high transmission redundancy. In our proposed coding scheme we construct a codeword which is not a linear combination of any $\gamma$ codewords previously transmitted to mitigate such transmission redundancy. We then note the observation that the probability of receiving a linearly dependent codeword is highest when the receiver has received $k-1$ linearly independent codewords. We propose using the blockACK frame so that the codeword received after $k-1$ linearly independent codeword is always linearly independent, this reduces the expected redundancy by a factor of three.
\end{abstract}

\section{Introduction} \label{sect:introduction}
Fountain codes are a class of forward error correction (FEC) coding scheme which corrects packet erasures independent of packet acknowledgement (ACK) or negative acknowledgement (NACK) information. Such erasure may be caused due to packet collision, packet drop due to congestion, and burst errors due to signal fading and channel noise in wireless channel. While the Automatic Repeat reQuest (ARQ) has been traditionally used to correct packet erasures in wireless network to improve reliability, collecting ACK frames incurs transmission and MAC overhead and hence reduces the throughput of the network. 

Fountain codes are used in a wide variety of transmission networks. The use of fountain coding scheme has shown to significantly outperform the hybrid ARQ (HARQ) scheme in 4G WiMAX network~\cite{Jin08}. It is projected that new modulation and coding schemes will be enabling technology to improve network capacity in 5G networks~\cite{Li14}. Fountain code based transmission protocol has been shown to outperform the transmission control protocol (TCP) to transmit flows in the Internet~\cite{Molnar14}. It has also been shown that the use of fountain code significantly improves the network throughput of a reprogramming protocol to disseminate code update to all sensors in a wireless sensor network~\cite{Rossi10}.

The fundamental idea of fountain code is that once the client has received $k(1+\epsilon)$ codewords of which at least $k$ are linearly independent, it can use Gaussian elimination to decode the codewords to recover the $k$ input packets. This saves the transmitter from collecting information about which codewords the client has successfully received and which ones are erased. The term $\epsilon$ is the overhead of the coding scheme. 

In this paper we propose a modified random linear fountain code (RLFC) scheme tailored for delay intolerant applications such as multimedia application which uses small packet generation size~\cite{Li16}, while taking into consideration the energy cost consideration of performing decoding. Such energy cost consideration is critical as increasing number of users are using battery constrained devices such as smartphones and tablets. In a Cisco whitepaper it has been projected that global mobile data traffic will increase eight-fold between 2015 and 2020, of which three fourth of the total traffic will be video~\cite{Cisco}. Small generation size is also used when coding is performed by dividing the packets of a file in to smaller sub-generations to minimize the energy cost of decoding even when the application is not delay intolerant~\cite{Sorensen14, Li11}.

RLFC suffer from high encoding and decoding complexities given as $\mathcal{O}(k^2L)$ and $\mathcal{O}(k^3+k^2L)$ respectively, where $L$ is the length of the packet, assuming multiplication tables are used to perform multiplication. For decoding, the term $k^3$ is the computational complexity of performing Gaussian elimination, and the term $k^2L$ is the computational complexity of multiplying the inverted matrix with the codewords. High encoding and decoding computational complexities in addition to incurring high energy cost, also contribute towards increased latency of decoding the input packets.

Coding over $GF(2)$ is in particular unique over other Galois field sizes because it does not require multiplication, as its field elements are given as $\{0,1\}$, this ensure that no more than half of the coding coefficients are non-zero. This in addition to saving the energy cost of multiplication table lookup, means that the number of XOR addition steps during encoding is given as $k^2L/2$. Similarly energy cost of table lookup is saved during Gaussian elimination and the number of XOR addition when multiplying the inverted matrix with codewords is given as $k^2L/2$.

Testbed implementations confirm these results. Implementation of RLFC on iPhone platform has shown that the encoding and decoding throughput for $GF(2)$ is 6-10 faster than $GF(2^8)$, depending on packet generation size given as $16\leq k\leq 256$, and packet length~\cite{Vingelmann10}. Similarly implementation of RLFC on TmoteSky sensor node has shown that decoding codewords generated over $GF(2)$ is at least 6.5 times faster than decoding codewords generated over $GF(2^8)$ for $16\leq k\leq 80$~\cite{Rossi10}.

Therefore despite the fact that the big O encoding and decoding computational complexities of $GF(2)$ based RLFC over $GF(q>2)$ based RLFC are same, the coding complexities of $GF(2)$ based RLFC are smaller than $GF(q>2)$ based RLFC by a large constant factor. Coding over smaller field size also reduces the header overhead given by $k \log_2 q$ bits to include the coding coefficient in the codeword's header.

However the use of small field size increases linear dependency of codewords, and leads to a higher decoding failure probability after the receiver has received $k+\delta$ codewords. It has been shown that $GF(2)$ based RLFC requires an average of around $k+1.6$ codewords before it can decode the $k$ input packets, whereas for $GF(2^8)$ based RLFC, an average of around $k+0.004$ codewords are sufficient~\cite{Cruces11}.

For applications where small $k$ is desirable such as in delay-intolerant transmission, multicast streaming (standardized by the IEEE 802.11aa Task Group), and wireless network with lossy transmission channel, an overhead of 1.6 codewords can adversely affect the network throughput. For instance, for packet batch size of 5 packets, a redundancy of 1.6 packets corresponds to an overhead of 32\%.

In this paper we first propose a modification to the RLFC scheme over smaller field size to reduce the decoding failure probability, by transmitting a codeword which is not given by linear combination of any $\gamma$ transmitted codeword. We then propose to leverage information from a single blockACK frame which reduces the expected number of codewords which a receiver need to receive before decoding the $k$ input packets from $k+1.6$ codewords to less than $k+0.6$ codewords. Mathematical models of the proposed modifications are presented which are verified with simulation result.

The rest of the paper is organized as follow. In Section~\ref{sec:preliminaries} we present the system model, notations used in the paper and bibliography of related work. In Section~\ref{sec:proposed} we then present our proposed modifications for RLFC and derive the mathematical model of its performance. Numerical results of the proposed scheme compared with the traditional RLFC scheme is given in Section~\ref{sec:results}. We then conclude with main result of our paper in Section~\ref{sec:conclusion}.

\section{Preliminaries}\label{sec:preliminaries}
\subsection{System Model and Notations}
We consider a transmitter transmitting $k$ packets, which we call input packets and denoted by $S=[s_1, s_2, \ldots, s_k]$, to $n$ clients. In this paper we consider both the unicast and multicast transmission. The network topology could be a wireless local area network (WLAN), or a multi-hop internet connecting server and clients across different networks. 

For abstraction and without loss of generality we consider a simple one-hop WLAN topology. Packet erasure at each client is assumed to be independent and identically distributed (iid), which follows the Bernoulli model with packet erasure probability of $p$, $0\leq p\leq 1$. The objective of the RLFC scheme is to transmit sufficient minimum codewords so that the receivers can decode the $k$ input packets. The term $(m,k)$ represents the number of codewords $m$ generated from $k$ input packets.

To generate a codeword, the transmitter first generates a coding coefficient vector $G_j=[g_1, g_2, \ldots, g_k]$. Each coding coefficient $g_i\in GF(q)$ is randomly and uniformly selected from the Galois field $GF(q)$ of size $q$. The generated coding coefficient vector is then multiplied with the input symbols to generate a codeword $c_j$  given as $c_j=G_j\cdot S^T$.

Once a receiver has collected $k$ linearly independent codewords, decoding is performed as $H^{-1}\cdot C^T$. Where the matrix $H$, $H\in GF(q)^{k\times k}$, represents the coding coefficient matrix of the $k$ linearly independent codewords a client has received, and the matrix $C$ represents the matrix of $k$ linearly independent codewords.

\subsection{Related Work}
Blasco and Liva have proposed a concatenated (15,10) RS code and RLNC $GF(16)$ generator matrix~\cite{Blasco11}. Their results have shown that by concatenating the RLNC with an MDS code such RS code, the decoding failure probability can be reduced by a factor of upto four assuming that the channel erasure rate are from low-moderate. However their work assumes that coding is performed over a non-binary Galois field $GF(q>2)$, and hence does not address the issue of high decoding computational complexity associated with non-binary coding coefficients.

Sorensen \textit{et al.} have proposed to use overlapping generations to minimize the decoding complexity~\cite{Sorensen14}. The general idea of performing encoding on overlapping generation is to divide the $k$ input packets into smaller sub-generations, each sub-generations having smaller number of input packets. By reducing the value of $k$ the decoding complexity is also minimized, however this approach adversely affect the network throughput~\cite{Li11}, and the use of overlapping sub-generations has been shown to improve throughput. In a multicast network this approach may lead to redundant codewords being received by some receivers, as few receivers are satisfied before others. 

Cruces \textit{et al.} have derived the exact decoding probability of decoding the codewords once a receiver has received $k+\delta$, $\delta\geq 0$, codewords~\cite{Cruces11}. Based on this result they derived the expected number of excess codewords a receiver needs to receive before performing decoding.

\section{Proposed Modification}\label{sec:proposed}
In our proposed modification to RLFC we randomly generate a coding coefficient vector, and then perform Gaussian elimination of this vector with all $\gamma$ combination from the $U$ transmitted codewords. For unicast transmission the expected number of transmissions before a receiver successfully receives a packet follows the geometric distribution, and the expected value of $U$ will be equal to $(1-p)^{-1}\cdot k$.

If the generated codeword is linearly independent with respect to all $\gamma$ combinations of coding vectors from $U$ then we generate the codeword and transmit the codeword. If the vector is not linearly independent with all the $\gamma$ combinations, then a new coding coefficient vector is randomly generated. As we consider small bounded values for $\gamma$, given as $\gamma\leq 3$, the computational complexity of performing Gaussian elimination on such small $\gamma\times k$ matrix is given as $\mathcal{O}(k)$. In the worst case scenario such Gaussian elimination may need to be performed for ${U\choose \gamma}$ times, leading to a total complexity of $\mathcal{O}(k^4)$ to generate one coding vector, as $U$ is given as a function of $k$ and assuming $\gamma\leq 3$. Such operation may need to be performed at least $k$ times, leading to a total complexity of $\mathcal{O}(k^5)$.

While the proposed modification leads to additional computation cost at the transmitter, we justify the higher computation cost at the transmitter, as the transmitter is usually a large device such as an access point, server, or a base transceiver station with high processing capabilities. While the receiver is usually a battery-constrained device such as a smartphone or laptop with limited processing capabilities. Our proposed modification assumes that the generation size is small and given as $k\leq 10$, and hence such additional computation cost at the encoder is not a cause of concern in practical networks.

Clearly our proposed coding scheme assumes that the following inequality is satisfied,

\begin{equation}\label{eq:inequality}
q^k - \sum_{i=0}^{\gamma} {U\choose i} (q-1)^i >0.
\end{equation}

\noindent When such an inequality is not satisfied the encoder can then generate coding vector randomly without checking for linear dependency of the generated coding vector with any $\gamma$ combinations from the transmitted codewords.

\begin{figure*}
\centering
\subfigure[$k=5$]{%
\centering
\includegraphics[width=0.48\textwidth]{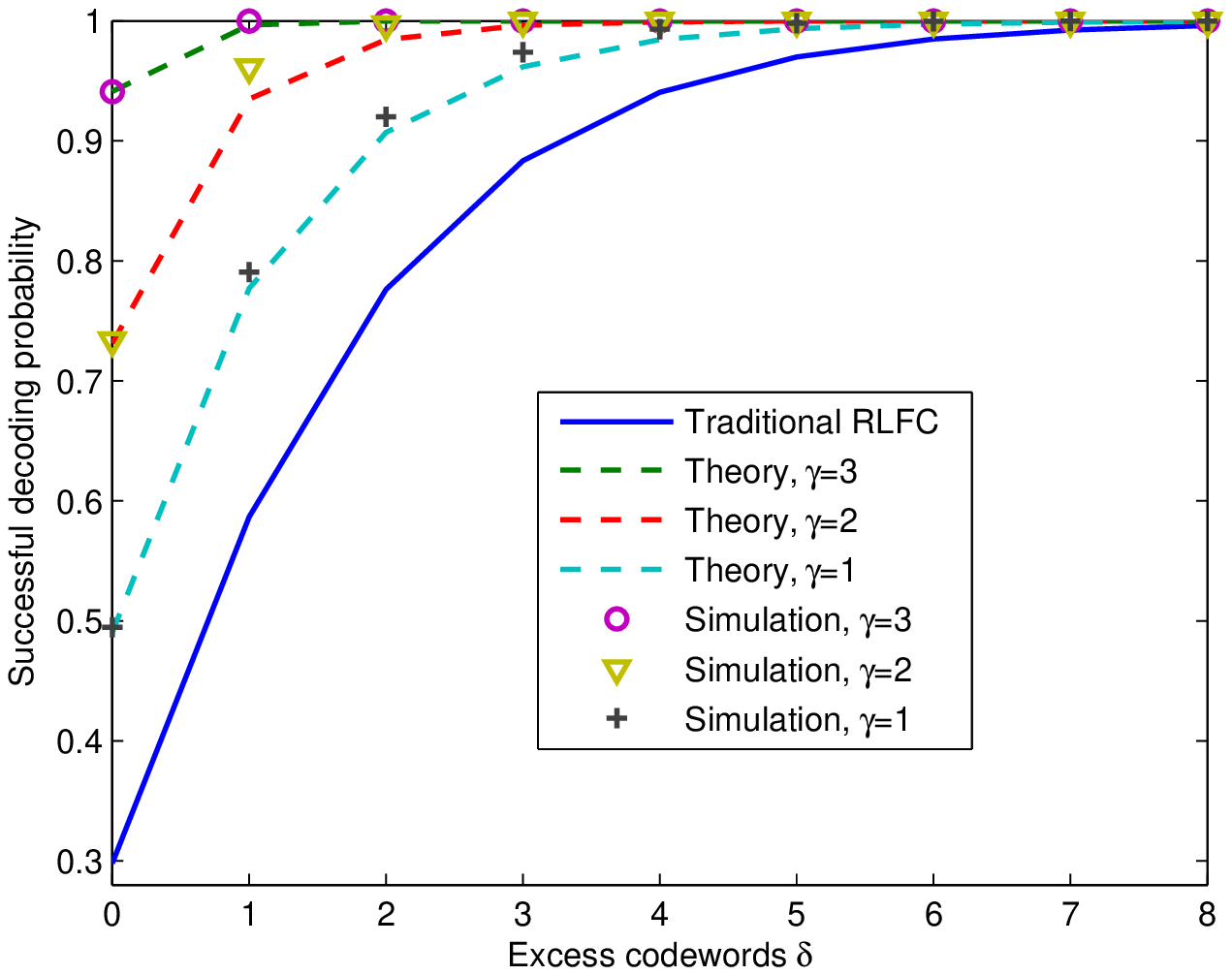}
\label{fig:subfigure1}}
\subfigure[$k=10$]{%
\centering
\includegraphics[width=0.48\textwidth]{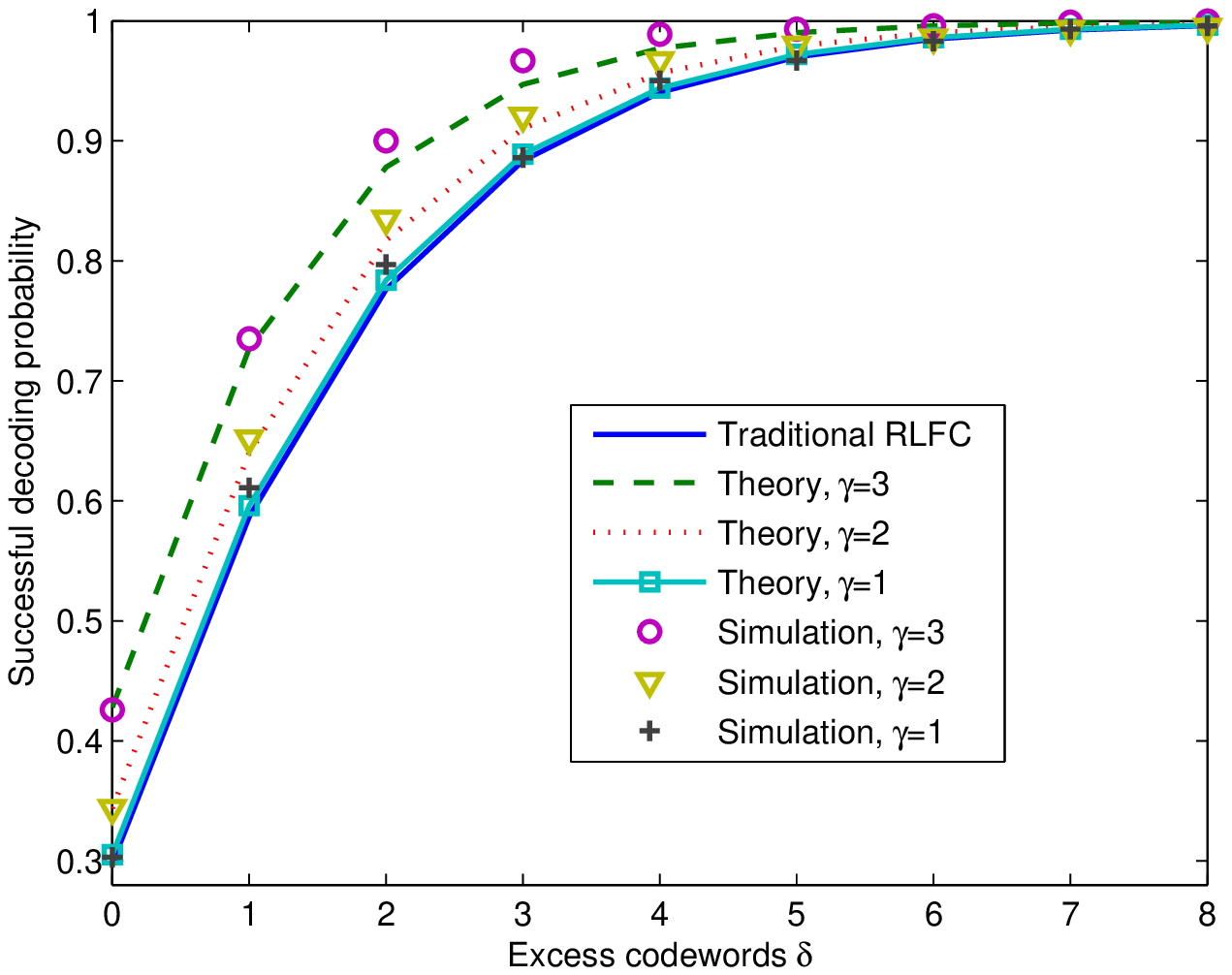}
\label{fig:subfigure2}}
\caption{Successful decoding probability for an excess codewords of $\delta\leq 8$, when (a) $k=5$, and (b) $k=10$, of the traditional RLFC and the modified RLFC over $GF(2)$.}
\label{fig:decode_prob}
\end{figure*}

\subsection{Performance Modeling}
We first present the mathematical model for the performance analysis of the proposed method under the idealistic assumption of lossless transmission channel. Let $H^k_\lambda$ denote the event that $H$ forms a full rank matrix after receiving exactly $\lambda$ codewords. And let $h^k_\delta$ denote the event that $H$ forms a full rank matrix after receiving $\delta$ excess codewords.

\textbf{CASE $\delta=0$.} As the proposed coding scheme guarantees that the first $\gamma+1$ codewords will not have a linearly dependent codeword, the probability that all the first $\gamma+1$ codewords are linearly independent is equal to one. The probability $P(D_{\gamma+1})$ that after receiving $\gamma+1$ codewords, the $\gamma+2^{nd}$ codeword will be linearly dependent is given as,
\[
P(D_{\gamma+1}) = \frac{q^{\gamma+1} - \sum_{i=0}^{\gamma} {\gamma\choose i} (q-1)^i} {q^k - \sum_{i=0}^{\gamma} {\gamma\choose i} (q-1)^i}.
\]

\noindent The term $q^{\gamma+1}$ represents the number of linearly dependent codewords which can be generated from the $\gamma+1$ transmissions, and the term after the minus sign represents the number of linearly dependent codewords which the encoder will not generate. Using the same argument, it can be shown that after any $u^{th}$ transmission, the probability that the $u+1^{th}$ codeword is linearly dependent is given as,

\[
P(D_u) = \frac{q^u - \sum_{i=0}^{\gamma} {\gamma\choose i} (q-1)^i} {q^k - \sum_{i=0}^{\gamma} {\gamma\choose i} (q-1)^i}.
\]

The probability that the $u+1^{th}$ codeword will be linearly independent is given as $1-P(D_u)$. It can now be shown that the probability that after receiving the $k^{th}$ codeword, $H$ forms a full rank matrix is given as follow,

\[
P(H^k_k) = \prod_{u=\gamma+1}^{k-1} 1-P(D_u).
\]

\textbf{CASE $\delta=1$}. The above scenario assumes that the excess codewords $\delta$ is equal to zero. We now derive the probability that the receiver need to receive exactly one excess codeword before $H$ forms a full rank matrix. In this case, it is only possible for one of the $\gamma+2^{nd}$ received codeword until the $k^{th}$ received codeword to be linearly dependent. The probability that the $\gamma+2^{nd}$ codeword is dependent is given by $P(D_{\gamma+1})$, the probability that the $\gamma+3^{rd}$ codeword is dependent is given by $P(D_{\gamma+2})$, and so on. Therefore the probability that a receiver needs to receive $k+1$ codewords is given by the summation of probabilities that one of the $\gamma+2^{nd}$ received codeword until the $k^{th}$ received codeword is linearly dependent multiplied by $P(H^k_k)$, as the remaining $k$ codewords form a full rank matrix.

\[
P(H^k_{k+1}) = P(H^k_k) \cdot \sum_{\mathclap{a_1=\gamma+1}}^{k-1} P(D_{a_1}).
\]

\textbf{CASE $\delta=2$}. We now extend our result for the case when the receiver receives two excess codewords before forming a full rank matrix. In such a case the first linearly dependent codeword can be the $\gamma+2^{nd}$ received codeword until the $k^{th}$ received codeword, corresponding to the events when the rank of $H$ is given as $\gamma+1$ until when the rank of $H$ is equal to $k-1$. Similarly the second linearly dependent codeword can be the $\gamma+3^{rd}$ received codeword until the $k+1^{th}$ received codeword, corresponding to the events when the rank of $H$ is given as $\gamma+1$ until when the rank of $H$ is equal to $k-1$. 

If $P(D_{a_1})$ denotes the probability that the first linearly dependent codeword is received when the rank of $H$ is $a_1$, $a_1\geq \gamma+1$. And $P(D_{a_2})$ denotes the probability the second linearly dependent codeword is received when the rank of $H$ is $a_2$, $a_2\geq \gamma+1$. Then the probability that exactly two excess codewords are required before $H$ form a full rank is given as,

\[
P(H^k_{k+2})=P(H^k_k)\, \cdot\, \sum_{\mathclap{\substack{\forall a_i\,:\,a_1\leq a_2, \\\gamma+1\leq a_i\leq k-1}}}\,\,\,\, \Big\{\prod\nolimits_{t=1}^2 P(D_{a_t})\Big\}. 
\]

\textbf{CASE arbitrary $\delta$}. Based on the above cases, for any arbitrary $\delta$, the probability that a receiver need to receive exactly $k+\delta$ codewords can be generalized as,
\[
P(H^k_{k+\delta})=P(H^k_k)\,\, \cdot\,\, \sum_{\mathclap{\substack{\forall a_i\,:\\\,a_1\leq a_2\leq\ldots\leq a_{\delta}, \\\gamma+1\leq a_i\leq k-1}}}\,\,\, \Big\{\prod\nolimits_{t=1}^{\delta} P(D_{a_t})\Big\}. 
\]

The probability $P(h^k_{\delta})$ that the receiver forms a full rank matrix $H$ if it has received $\delta$ excess codewords is given as,

\[
P(h^k_{\delta})=\sum_{i=0}^{\delta} P(H^k_{k+i}).
\]

\noindent And the expected number of excess codewords which a receiver needs to receive before decoding the $k$ input packets is given as,

\[
E[\delta] =\sum_{\delta=0}^{\infty} (k+\delta)\cdot P(H^k_{k+\delta})
\]

\begin{figure*}
\centering
\subfigure[$p=0.05$]{%
\centering
\includegraphics[width=0.48\textwidth]{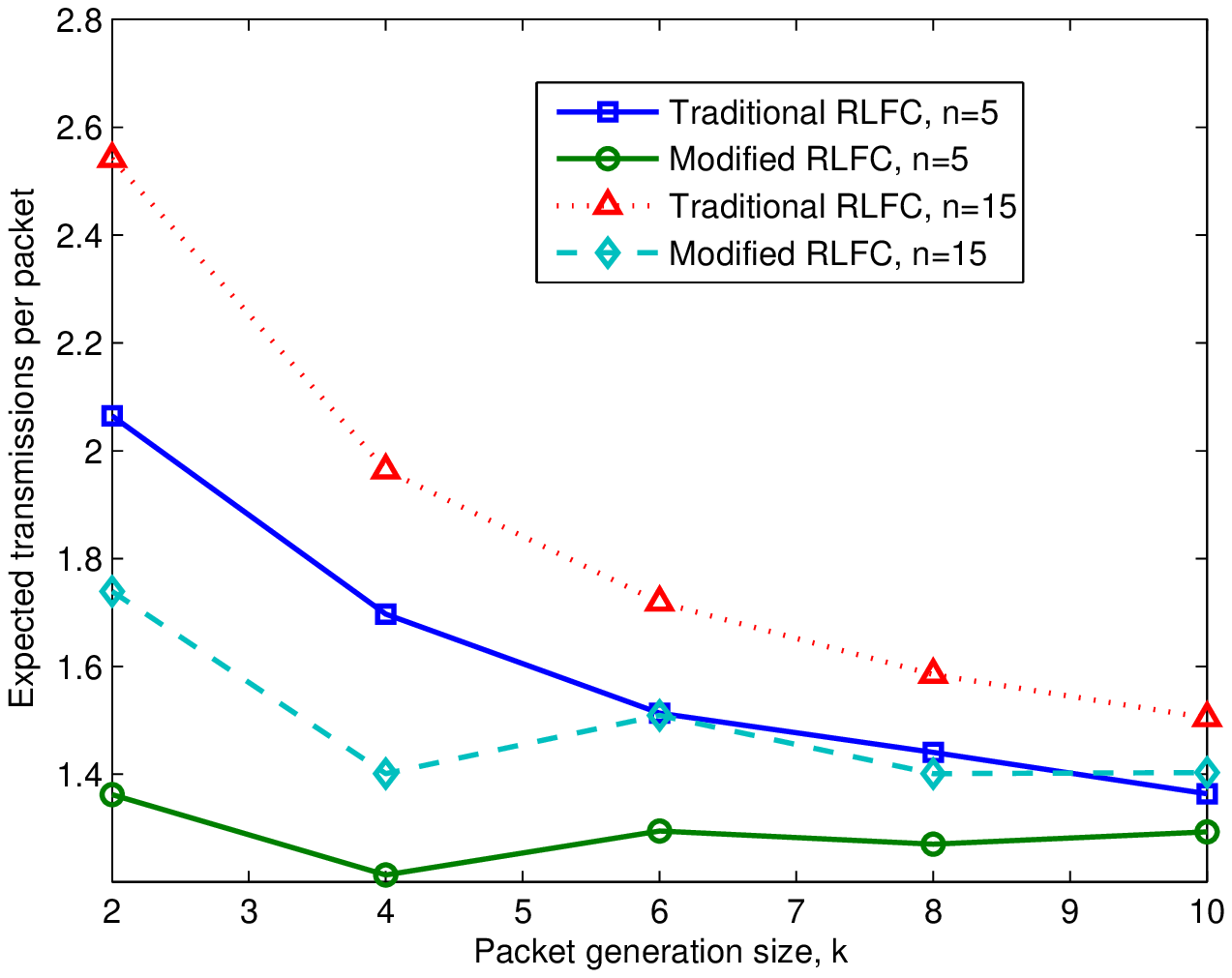}
\label{fig:subfigure1}}
\subfigure[$p=0.2$]{%
\centering
\includegraphics[width=0.48\textwidth]{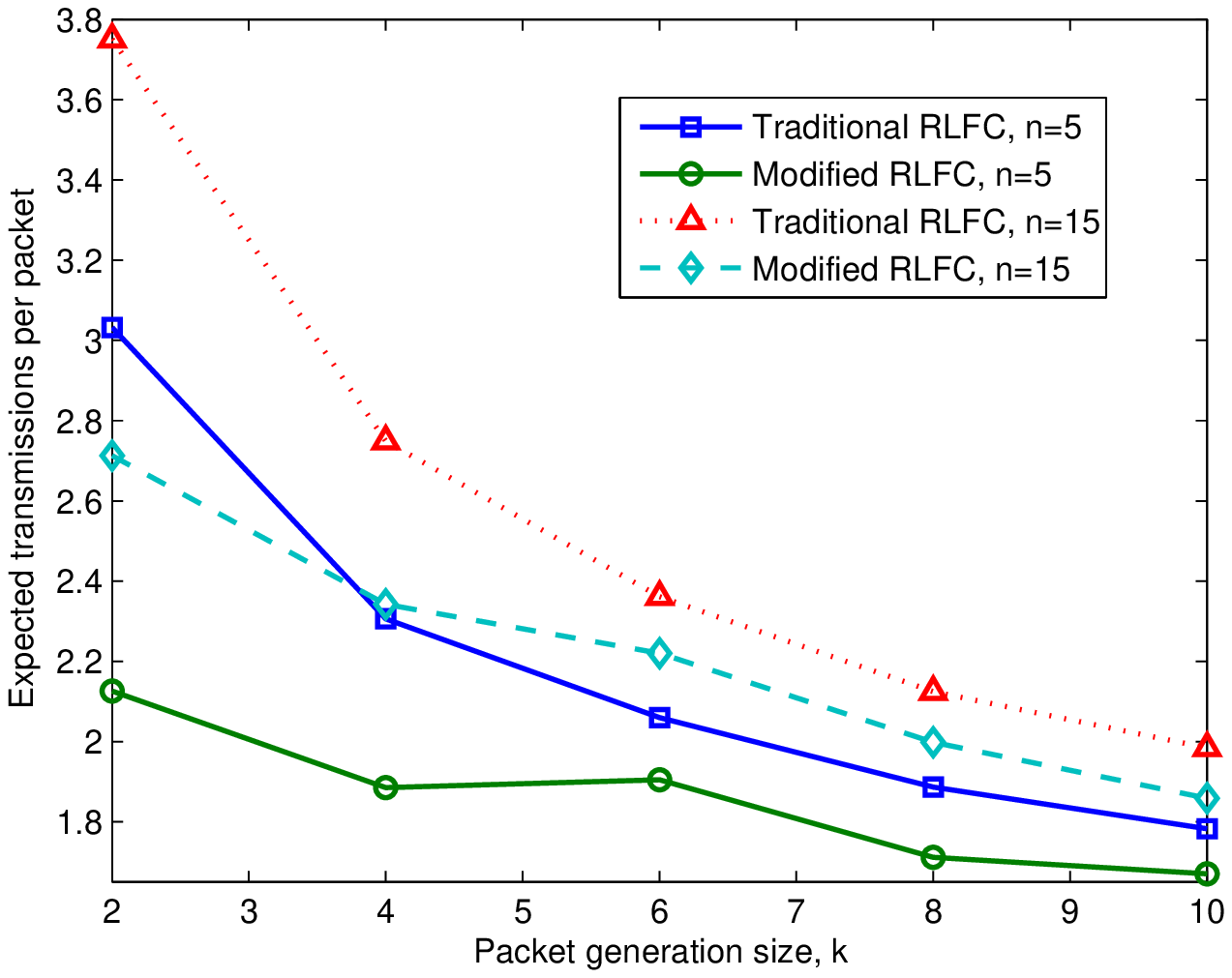}
\label{fig:subfigure2}}
\caption{Expected number of transmissions of the modified RLFC compared with the traditional RLFC using simulation for a single-hop multicast network for (a) $p=0.05$, and (b) $p=0.2$.}
\label{fig:multicast}
\end{figure*}

\begin{figure*}
\centering
\subfigure[Without the use of blockACK frame]{%
\centering
\includegraphics[width=0.48\textwidth]{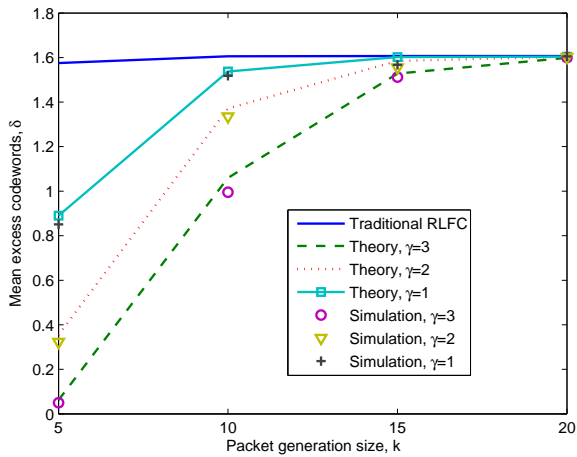}
\label{fig:subfigure1}}
\subfigure[With the use of blockACK frame]{%
\centering
\includegraphics[width=0.48\textwidth]{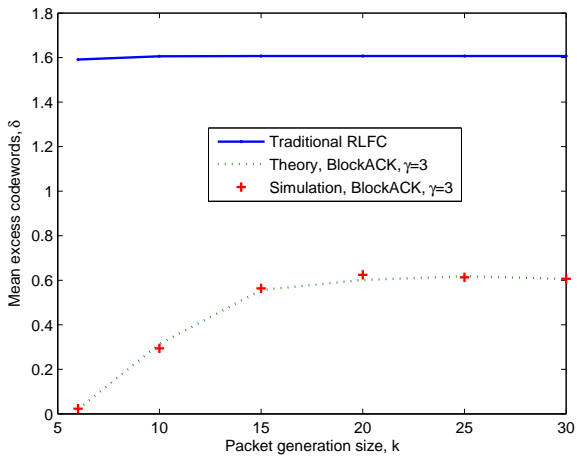}
\label{fig:subfigure2}}
\caption{The expected number of excess codewords required for the traditional RLFC compared with the (a) modified RLFC without using the blockACK frame, and (b) when using the blockACK frame.}
\label{fig:excesspks}
\end{figure*}

\subsection{BlockACK based Modification}
For the modification proposed in the previous section, an improvement in $E[\delta]$ is limited for smaller generation size. To reduce the excess codewords for larger generation size, we propose to use the BlockACK. The use of the BlockACK frame is motivated by the fact that the probability that a receiver receives a linearly dependent codeword is highest when it has received $k-1$ linearly independent codewords. In our proposed scheme, the receiver transmits a BlockACK frame which include the coding coefficients of all the $k-1$ linearly independent codewords. The transmitter uses this information to construct a codeword which is linearly independent for the receiver. 

Therefore in this paper we apply the BlockACK based modification for the simple case of unicast transmission. When such a modification is applied, the probability $P(H^k_{k, BK})$ that the receiver needs zero excess codeword is given as,

\[
P(H^k_k, BK) = \prod_{u=\gamma+1}^{k-2} 1-P(D_u).
\]

\noindent And the probability that it needs an arbitrary excess packets $\delta$ is given as,

\[
P(H^k_{k+\delta, BK})=P(H^k_{k, BK})\,\, \cdot\,\, \sum_{\mathclap{\substack{\forall a_i\,:\\\,a_1\leq a_2\leq\ldots\leq a_{\delta}, \\\gamma+1\leq a_i\leq k-2}}}\,\,\, \Big\{\prod\nolimits_{t=1}^{\delta} P(D_{a_t})\Big\}. 
\]

\noindent The expected number of codewords which the receiver now needs to receive is given as,

\[
E[\delta_{BA}] =\sum_{\delta=0}^{\infty} (k+\delta)\cdot P(H^k_{k+\delta, BK}).
\]

\section{Numerical Results}\label{sec:results}
In this section we compare the performance of our proposed modified RLFC with the traditional RLFC over $GF(2)$ using mathematical models proposed in the previous section, which is verified using simulation results. The results for the traditional RLFC are constructed from the model proposed in~\cite{Cruces11}.

The decoding success probability of the modified RLFC compared with the traditional RLFC is plotted in Figure~\ref{fig:decode_prob}. The figure shows that the decoding success probability increases with $\gamma$ and is higher for small values of $k$. The theoretical result matches fairly well with the simulation result. For each set of parameters 500 simulation runs were performed. The results in this figure are plotted based on the assumption that the transmission channel is lossless and hence inequality~\eqref{eq:inequality} is satisfied. While this assumption may not be valid in practical networks, it provides an upper bound on the decoding success probability.

To take into consideration the more practical scenario of lossy transmission channel, we conduct simulation to compare the performance of the traditional RLFC and modified RLFC for multicast network which is plotted in Figure~\ref{fig:multicast}. The results show that for transmission channel with low-moderate erasure probability, the modified RLFC can reduce the number of transmissions due to its higher decoding success probability.

In Figure~\ref{fig:excesspks}, the expected number of excess codewords $E[\delta]$ which a receiver need to receive before successfully decoding the $k$ input packets is plotted. For small packet generation size, the modified RLFC can significantly reduce $E[\delta]$. Without the use of blockACK, the value of $E[\delta]$ converges to 1.6 for generation size of 20 packets. 

However with the aid of the blockACK frame the value of $E[\delta]$ can be significantly reduced as shown in the figure. From the figure it can be seen that such an approach can reduce $E[\delta]$ from 1.6 to 0.6. We have in fact benefitted in reduction of one redundant codeword whose length can be upto 1-2KB, by transmitting a blockACK whose length will be approximately less than 0.1KB for $k\leq 30$.

\section{Conclusion}\label{sec:conclusion}
In this paper we studied the problem of increasing the decoding success probability of RLFC over $GF(2)$. Such codes enjoy low decoding complexity compared to RLFC over larger field size, and low header overhead. However due to the use of small field size, they suffer from low decoding success probability.

We proposed a modification to the RLFC code, by transmitting only those codewords which are not linear combination of any $\gamma$ codewords previously transmitted. This modification has shown potential to noticeably reduce the expected number of transmissions in lossy multicast network for packet generation size of $k\leq 10$. Such small packet generation size are used in delay-intolerant multimedia transmission scheme to provide user high quality of service (QoS), and in coding schemes which uses smaller sub-generation to minimize the decoding computational complexity. 

With an increase in mobile data traffic, especially of real-time traffic such as video streaming, the proposed coding scheme increases the throughput in such networks while keeping the energy cost consumption minimized. Energy cost consideration is critical in battery constrained devices such as smartphones.

Analytical model of the proposed modified RLFC was proposed, and simulation results on the performance of such coding scheme for a multicast network with low-moderate packet erasure probability presented. Simulation results show that the proposed coding scheme can significantly reduce the expected number of transmissions for a multicast networks.

We then proposed to take advantage of blockACK frame which is much smaller in length than the length of a data packet to reduce the number of redundant transmission from 1.6 to approximately 0.6. An interesting future work is to study and analyze how the proposed coding scheme can be integrated in existing overlapping generation based coding to improve the transmission throughput.

\bibliographystyle{IEEEtranS}
\bibliography{IEEEabrv,algorithm}

\end{document}